# Calculation and properties of trap structural functions for various spatially correlated systems

Arkadiusz Mandowski

*Institute of Physics, Jan Długosz University, ul. Armii Krajowej 13/15, PL-42-200 Częstochowa, Poland*
*e-mail: a.mandowski@ajd.czest.pl*

Thermoluminescence (TL) kinetics in spatially inhomogeneous systems can be studied by various Monte Carlo algorithms. Recently, a new analytical approach was suggested for the isolated cluster model. The theory is based on the concept of trap structural functions (TSFs). TSFs depend solely on topological properties of solids. Therefore, knowing TSFs for traps and recombination centres it is possible to calculate TL for various parameters, e.g. different heating schemes and different energy configurations. This paper presents some properties and methods of calculation of TSFs. Structural character of TSFs is verified numerically. It is shown that for simple cluster systems it is possible to calculate the functions analytically.

## I. INTRODUCTION

Classical theories of trapping and recombination of charge carriers in dielectrics relate to two analytically described cases. The first one relates to uniform distribution of traps and recombination centres (RCs), where transitions of charge carriers go through the conduction band. This is the simple trap model (STM)[1]. The second one relates to pairs of traps and RCs placed close to each other. The model of localized transitions (LT) assumes that recombination of trapped carriers proceeds through a local excited level[2,3]. Recent analysis of Townsend and Rowlands[4] gave an evidence that TL kinetic processes proceed in large scale defects. Obviously, no one of the above models (LT and STM) is able to describe this type of kinetics properly.

The studies of TL kinetics in various spatially correlated systems (SCSs) were started some years ago by Mandowski and Świątek[5]. For this purpose they used several Monte Carlo algorithms modelling TL e.g. in 1-D and 3-D systems under different external conditions. It was found that SCS TL shows many unexpected features that cannot be explained within the framework of LT and STM models. Examples include apparently composite structure of monoenergetic peaks[6,7], additional 'displacement' peaks[8] and the dependence of TL on the external electric field[9,10]. For review, see the papers[11,12]. Recently[13], an analytical model was proposed for the isolated clusters (IC) model which is a special case of SCSs. The theory is based on two trap structural functions (TSFs) for electrons and holes - $\Gamma_n$ and $\Gamma_h$, respectively. These functions depend only on structural properties (e.g. spatial distribution, energy barriers, etc.) of metastable states in a solid. This paper presents some properties and practical methods of calculation of TSFs.

## II. BASIC EQUATIONS

Commonly accepted explanation of long-lasting phosphorescence and TL phenomena is based on the assumption of metastable levels (traps and recombination centres) situated within the energy gap. Although direct transition from trap to a recombination centre is possible, most of the transitions takes place through excited states. This is shown schematically in the case of 'active' electron traps in Figure 1. This diagram relates also to the case of a set of IC, which are separated by a distance and/or energy barriers. Each cluster, containing 'active' traps, deep traps and RCs, has the energy configuration shown in Figure 1.

We allow the clusters to have various size (i.e. different number of traps and RCs). However, all trapping and recombination processes must occur only within the clusters, i.e. no interaction between clusters is possible. TL is produced in the same way as in the LT model - a trapped charge carrier is thermally excited to a local excited level and then it may be retrapped or may recombine with an opposite charge carrier trapped at the RC level. For the simplest case of a single type of active traps, deep traps and RCs the following set of kinetic equations was proposed[13]:

$$-\dot{n} = n\nu \exp\left(\frac{-E}{kT(t)}\right) - \Gamma_n(n)n_e, \tag{1a}$$

$$-\dot{h} = \Gamma_h(h)n_e, \tag{1b}$$

$$h = n + n_e + M, \tag{1c}$$

where $E$ stands for the activation energy and $\nu$ is the frequency factor for active traps. $n$, $n_e$ and $h$ denote the total concentrations of electrons trapped in active traps, electrons in the excited levels and holes trapped in RCs. $M$ stands for the concentration of electrons in the thermally disconnected traps (deep traps), i.e. traps that are not emptied during the experiment. $\Gamma_n$ and $\Gamma_h$ denote two TSFs for trapping and recombination respectively. For the two extreme cases - LT and STM, TSFs are the following: $\Gamma_h^{(\mathrm{LT})} = \bar{B}$, $\Gamma_n^{(\mathrm{LT})} = \bar{A}$, $\Gamma_h^{(\mathrm{STM})} = Bh$ and $\Gamma_n^{(\mathrm{STM})} = A(N-n)$, where $A$, $B$, $\bar{A}$ and $\bar{B}$ are constants and $N$ is the total concentration of traps (in the case of STM $n_e$ has the meaning of the concentration of carriers in the conduction band). Obviously, these functions do not depend e.g. on the heating rate $\beta$ and the activation energy of traps $E$. The $\Gamma$-functions are structural in the sense that these are unique for a system of traps and RCs characterized by a specific spatial distribution with definite transition probabilities. Nevertheless, this useful property is not self-evident for complex IC systems. Therefore it needs to be clarified to what extent the conclusions may applied to spatially correlated systems.

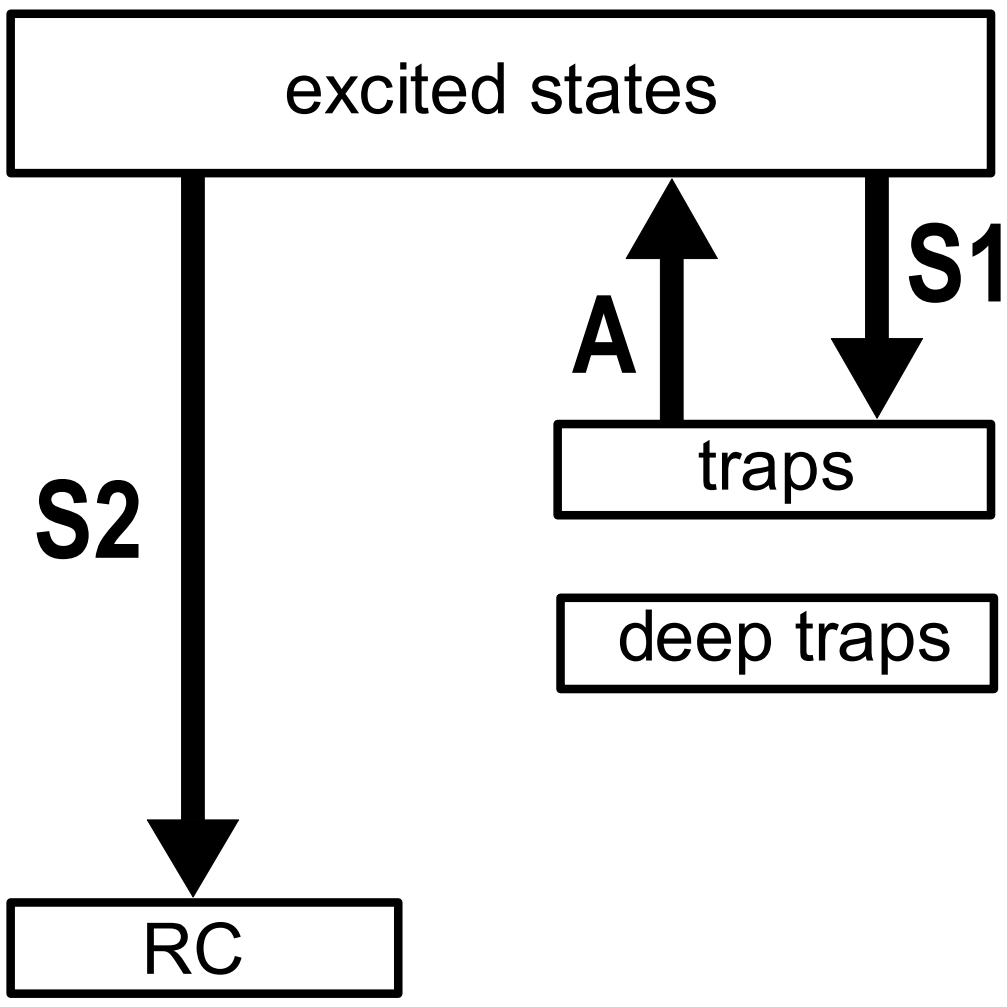

Figure 1. Energy diagram for a single cluster in the IC model, consisting of a single trap level, one kind of RCs and a number of deeper traps. The whole system consists of a very large number of clusters having the same energy configuration. The number of traps and RC in each cluster may be different.

## III. CALCULATION OF TSFs

### A. Direct numerical calculation - arguments confirming the invariance property

Monte Carlo algorithms for the IC model are performed by considering elementary transitions - $\Im_D$, $\Im_T$ and $\Im_R$ for detrapping, trapping and recombination, respectively:

$$\Im_D(t) = \nu \exp\left[\frac{-E}{kT(t)}\right] \tag{2a}$$

$$\Im_T(t) = \bar{A}\left[\bar{N} - \bar{n}(t)\right] \tag{2b}$$

$$\Im_R(t) = \bar{B}\bar{h}(t) \tag{2c}$$

Here, the dashed values denote variables and parameters relating to a single cluster of the system. $\bar{N}$ denotes the number of trap levels, $\bar{n}_e$ is the number of electrons in the local excited level, $\bar{n}$ is the number of electrons in traps and $\bar{h}$ denotes the number of holes in RCs. $\bar{A}$ and $\bar{B}$ denote coefficients for trapping and recombination, respectively. Detailed method of the simulation as well as the scaling properties defining relation between the microscopic and the macroscopic parameters were given in some previous papers(5,9).

For a given IC system TSFs can be calculated by performing Monte Carlo simulation and then calculating $\Gamma_n$ and $\Gamma_h$ from eqs (1a) and (1b), respectively. Some general properties of TSFs were presented in earlier paper(13). The key point is to test the invariance of TSFs with respect to thermal treatment. This obviously holds true for LT and STM models. From earlier Monte Carlo simulations we learned(6,7,9) that the greatest discrepancies in TL kinetics between the IC and the two standard models take place for small clusters ($\bar{n}_0 < 10$) characterized by high retrapping coefficients ($r \equiv \bar{A}/\bar{B} > 1$) and low density of thermally disconnected traps ($\omega = M/N < 1$). For that reason in this region one may expect also possible divergence of TSFs e.g. for various heating rates.

In Figure 2 and Figure 3 two TSFs are calculated for two cluster systems: $\bar{n}_0 = 2$ and $\bar{n}_0 = 5$. The simulations were performed assuming typical parameters: $E = 0.9\,\mathrm{eV}$, $\nu = 10^{10}\,\mathrm{s}^{-1}$, $r = 100$ and $\omega = 0$ for the heating rate $\beta$ ranging from $10^{-2}$ K/s to $10^2$ K/s. Additionally, $\Gamma_n$ and $\Gamma_h$ were calculated for the isothermal decay (phosphorescence) at 450 K. Taking into account natural statistical fluctuations inherent in Monte Carlo simulations one may notice excellent agreement between those curves. This result clearly confirms the invariance of TSF with respect to thermal treatment.

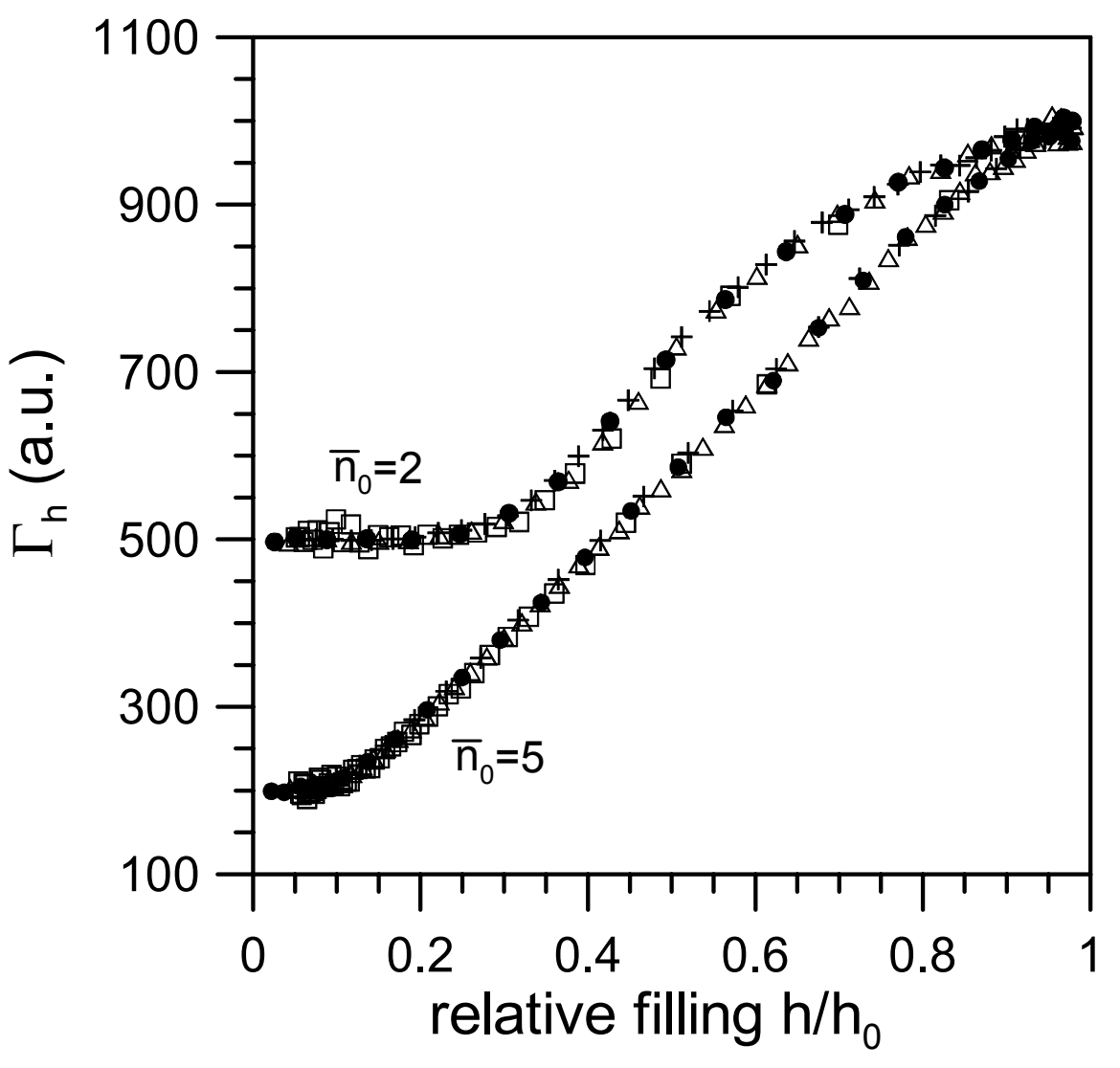

Figure 2. The function $\Gamma_h(h)$ calculated for two IC systems consisting of $\bar{n}_0 = 2$ and $\bar{n}_0 = 5$ charge carriers in a single cluster. The functions were computed using Monte Carlo simulation for various heating rates: $\beta = 10^{-2}$ K/s (+), $\beta = 1$ K/s (●) and $\beta = 10^2$ K/s (△). The squares (□) denote results obtained for isothermal decay (phosphorescence, $\beta = 0$) at $T = 450\,\mathrm{K}$. Other trap parameters: $E = 0.9\,\mathrm{eV}$, $\nu = 10^{10}\,\mathrm{s}^{-1}$ $r = 100$ and $\omega = 0$.

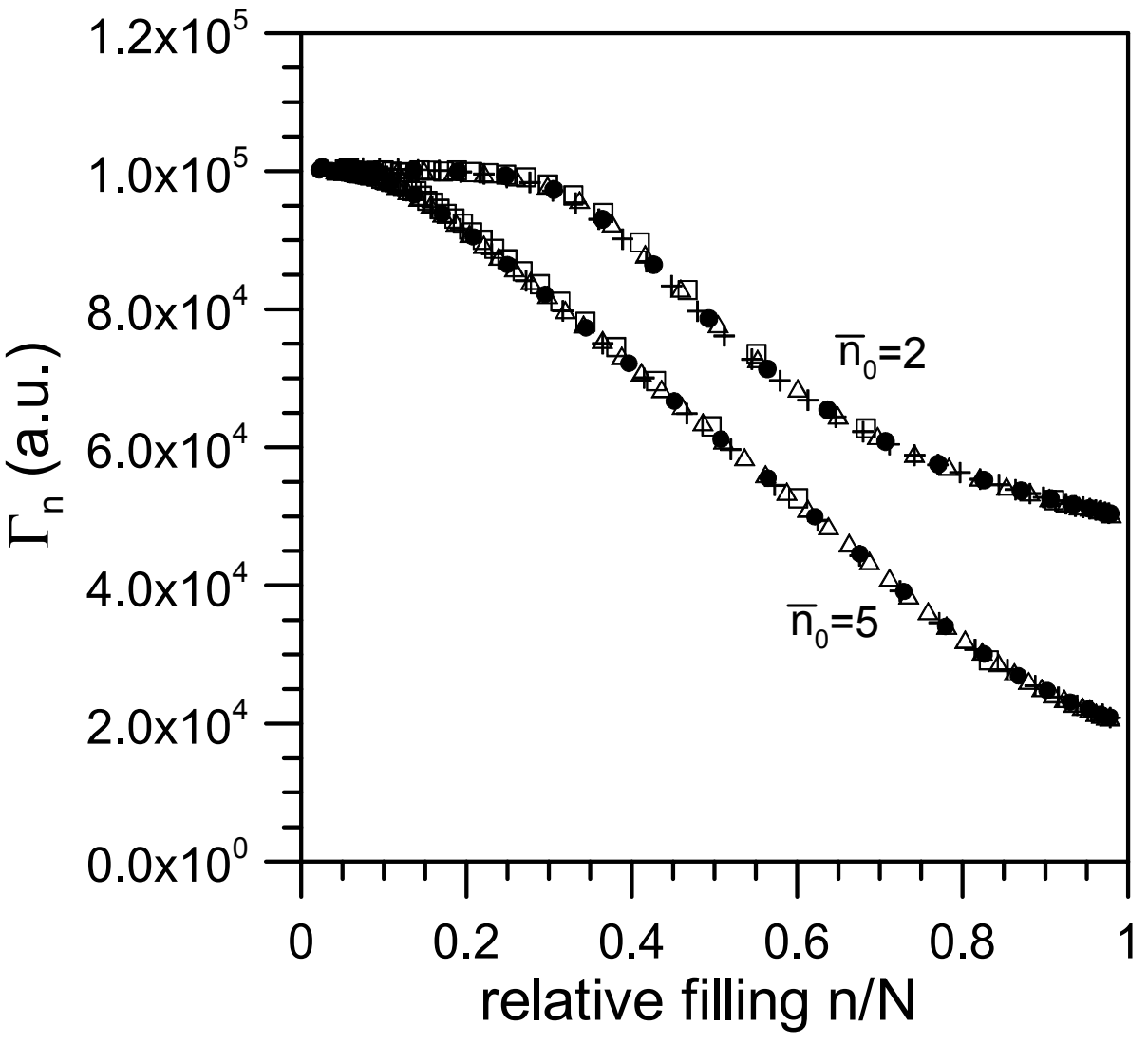

Figure 3. The function $\Gamma_n(n)$ calculated for two IC systems consisting of $\bar{n}_0 = 2$ and $\bar{n}_0 = 5$ charge carriers in a single cluster. All parameters and symbols the same as in Figure 2.

Another interesting feature is shown in Figures 4 and 5. The TSFs were calculated here for three different activation energies: $E = 0.9\,\mathrm{eV}$, $E = 1.1\,\mathrm{eV}$ and $E = 1.3\,\mathrm{eV}$. The retrapping coefficient was $r = 1$ and other parameters were the same as for previous cases. Such diverse activation energies yield transition probabilities which differ by many orders of magnitude. Nevertheless, the coincidence between $\Gamma$'s is excellent. This result supports the argument that TSFs do not depend also on the A transition (see Figure 1). Naturally, the transition is necessary to supply charge carriers to the excited state, however the time (and temperature) dependence is not essential.

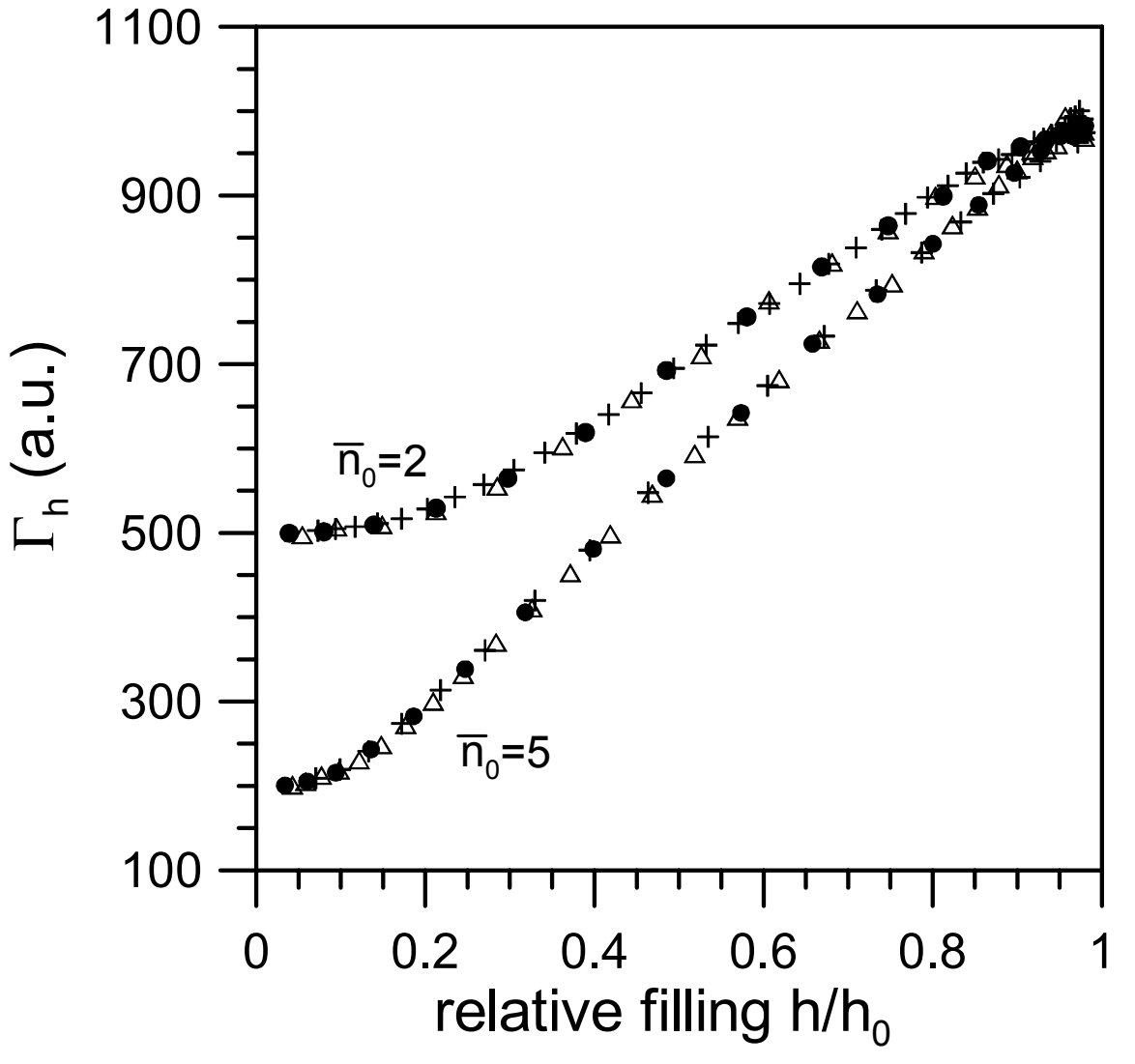

Figure 4. The function $\Gamma_h(h)$ calculated for two IC systems consisting of $\bar{n}_0 = 2$ and $\bar{n}_0 = 5$ charge carriers in a single cluster. The functions were computed for the retrapping coefficient $r = 1$ and various activation energies: $E = 0.9$eV (+), $E = 1.1$eV (●) and $E = 1.3$eV (△). Other parameters and symbols the same as in Figure 2.

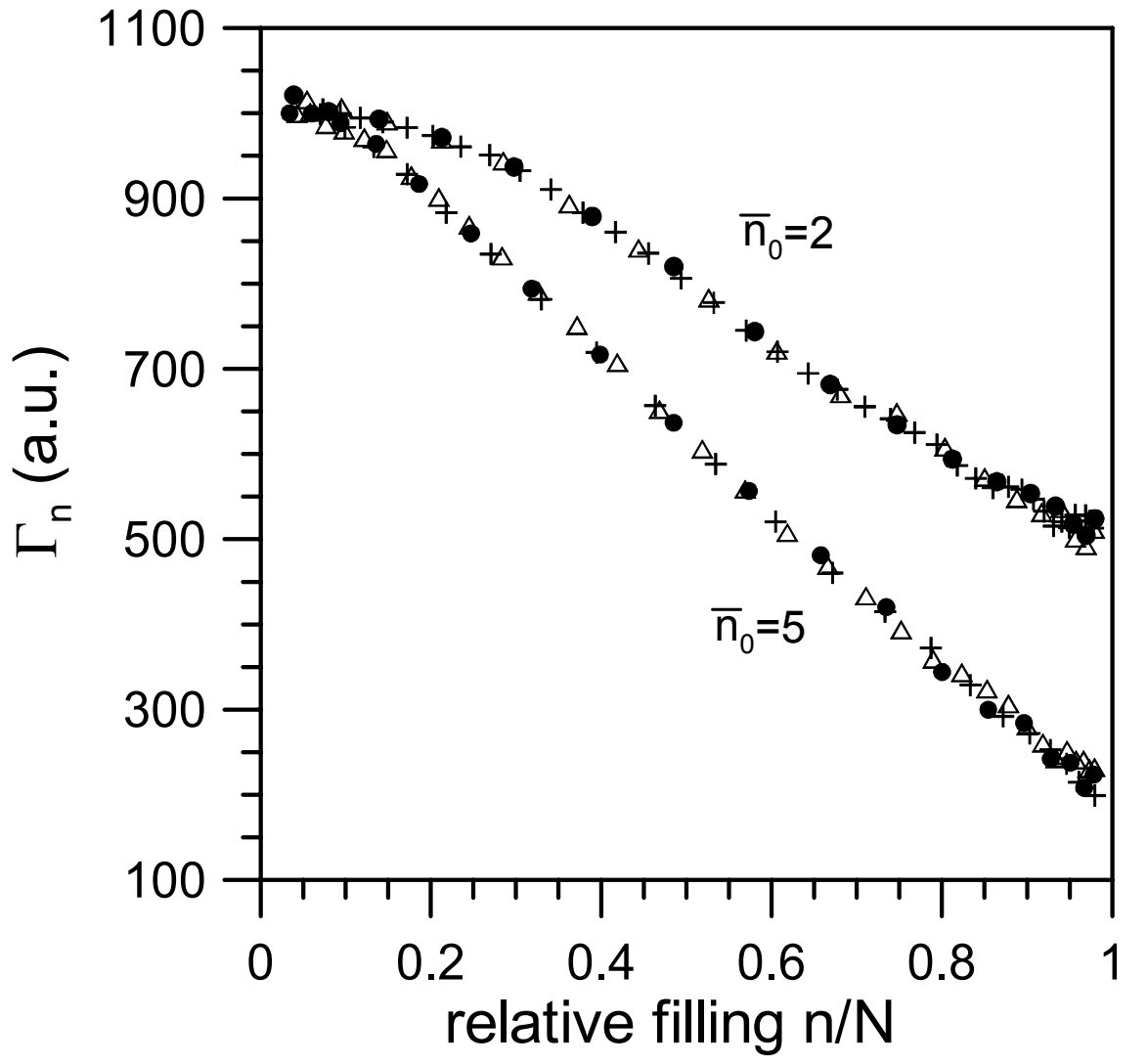

Figure 5. The function $\Gamma_n(n)$ calculated for two IC systems consisting of $\bar{n}_0 = 2$ and $\bar{n}_0 = 5$ charge carriers in a single cluster. All parameters and symbols the same as in Figure 4.

### B. Effective numerical calculation

A major disadvantage of the Monte Carlo simulation method is a very high computation time that is required for obtaining accurate kinetic data. This is especially true for systems characterized by high recombination coefficients. Another difficulty is related to the calculation of the probability density function corresponding to the detrapping probability given by eq. (2a). An integral appearing in the probability density function could not be solved analytically. To solve the integrals it is necessary to use some special time-consuming algorithms[5]. During typical Monte Carlo simulation the integrals have to be calculated at least $10^6...10^8$ times. Therefore, its efficient calculation is essential for the overall simulation performance.

Taking advantage of the invariance properties illustrated in Figures 2-5 it is possible to simplify considerably Monte Carlo calculations for the purpose of calculating TSFs. First of all it is possible to assume $\Im_D(t) = const.$ In this case the integral required for the calculation of the probability density function relating to $\Im_D$ becomes solvable. The same can be achieved by performing isothermal simulation as shown in Figures 2,3. This way calculated TSFs may be used for calculating TL in equivalent IC systems by solving numerically the set of eqs. (1).

### C. Results for various r

In Figure 6 the function $\Gamma_h$ is presented for two IC systems $\bar{n}_0 = 2$ and $\bar{n}_0 = 5$ calculated for various retrapping coefficients *r*. The variation of *r* was achieved by setting $\bar{B} = const.$ and changing only the $\bar{A}$ coefficient. For that reason it seems astonishing that so significant discrepancies occur for corresponding $\Gamma_h$ functions, especially for the smallest clusters with $\bar{n}_0 = 2$ . A similar dependence can be noticed for the $\Gamma_n$ . Below, we will give a qualitative explanation of this feature limiting ourselves to the case of $\bar{n}_0 = 2$ .

Using eq. (1b) we can define the $\Gamma_h$ function as:

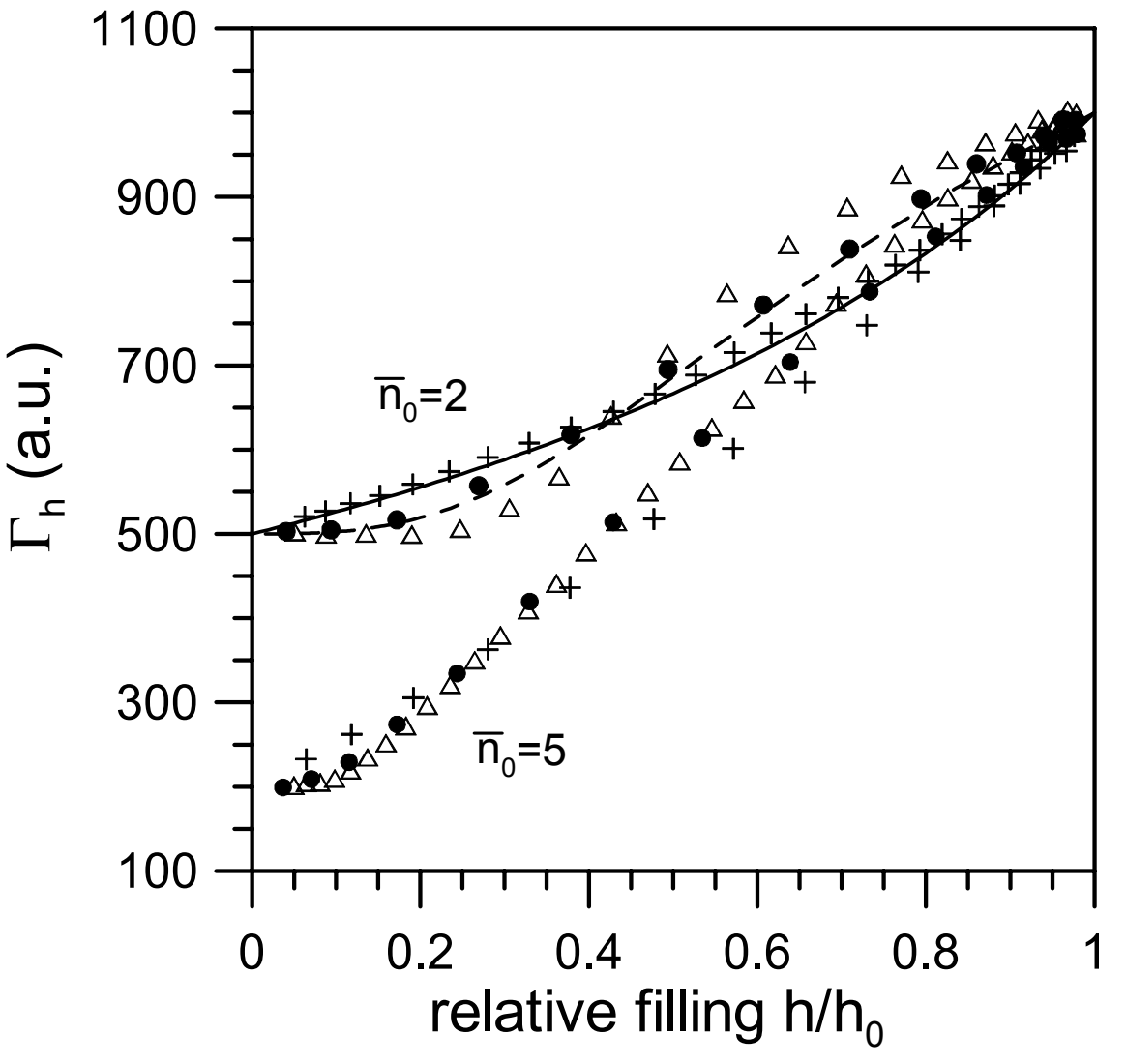

Figure 6. The function $\Gamma_h(h)$ calculated for two IC systems consisting of $\bar{n}_0 = 2$ and $\bar{n}_0 = 5$ charge carriers in a single cluster. The functions were computed for various the retrapping coefficients: $r = 0.01$ (+), $r = 1$ (●) and $r = 100$ (△). Other parameters are the same as in Figure 2. The lines denote $\Gamma_h(h)$ calculated theoretically. The solid line corresponds to the eq. (12) and the dashed line corresponds to the eq. (18).

$$\Gamma_h(h) = -\dot{h}/n_e \quad (3)$$

For a given number of holes in RCs $h(t)$ and a given number of electrons supplied to the excited states $n_e$ it is clear that the derivative $\dot{h}$ depends not only on $n_e$ but also on the distribution of excited charge carriers in all partially filled clusters. As the A transition (detrapping) has no influence on TSFs, the distribution will be chiefly determined by the retrapping ratio $r \equiv \bar{A}/\bar{B}$. Consequently, the retrapping coefficient must be considered as a parameter characterizing structural properties of the IC system.

### D. Analytical calculations

For its simplicity, the IC system with $\bar{n}_0 = 2$ allows analytical calculation of TSFs. We will consider two simple cases which are not directly related to any type of real TL kinetics, however they allow for easy qualitative description of TSFs. In the examples we will calculate the $\Gamma_h$ function. Let the variables $W_1(t)$ and $W_2(t)$ denote the concentrations of clusters (with two traps and two RCs $\bar{n}_0 = 2$ ) having one and two active charge carriers, respectively. Initially $W_1(0) = 0$ and $W_2(0) = h_0/2$, where $h_0$ is the initial concentration of holes.

First, let us assume that each cluster has only one electron in the excited level. In the case of recombination the electron is immediately replaced by the second one. For these conditions we can write two differential equations:

$$\dot{W}_1(t) = 2\bar{B}W_2(t) - BW_1(t) \quad (4)$$

$$\dot{W}_2(t) = -2\bar{B}W_2(t) \quad (5)$$

The calculation of $\Gamma_h$ requires previous calculation of $h(t)$ and $n_e(t)$. Here, the variables are determined as follows:

$$h(t) = W(t)_1 + 2W_2(t) \quad (6)$$

$$n_e(t) = W_1(t) + W_2(t) \quad (7)$$

Solving eq. (5) we get

$$W_2(t) = \frac{h_0}{2} e^{-2\bar{B}t} \quad (8)$$

Substituting the result to eq. (4) and solving the equation with respect to $W_1(t)$ leads to the following formulae:

$$W_1(t) = h_0\left(e^{-\bar{B}t} - e^{-2\bar{B}t}\right) \quad (9)$$

and

$$h(t) = h_0 e^{-\bar{B}t} \quad (10)$$

$$n_e(t) = h_0\left(e^{-\bar{B}t} - \frac{1}{2}e^{-2\bar{B}t}\right) \quad (11)$$

Finally, we get the following equation for $\Gamma_h$ :

$$\Gamma_h(h) = \frac{2\bar{B}}{2 - h/h_0} \quad (12)$$

This dependence is shown in Figure 6 as a solid line. The solution is very close to the case representing very low retrapping coefficient ( $r = 0.01$ ).

Now, let us consider another case. Initially all active charge carriers are in the excited level. Neglecting transitions A and S1 (Figure 1) we get the following set of equations:

$$\dot{W}_1(t) = 4\bar{B}W_2(t) - BW_1(t) \quad (13)$$

$$\dot{W}_2(t) = -4\bar{B}W_2(t) \quad (14)$$

The solution for $W_1(t)$ and $W_2(t)$ is:

$$W_1(t) = \frac{2h_0}{3}\left(e^{-\bar{B}t} - e^{-4\bar{B}t}\right) \quad (15)$$

$$W_2(t) = \frac{h_0}{2} e^{-4\bar{B}t} \quad (16)$$

The definition of $h(t)$ remains unchanged, but $n_e(t)$ is now:

$$n_e(t) = h(t) = W(t)_1 + 2W_2(t) \quad (17)$$

Using again the definition (3) we get finally parametric equation for $\Gamma_h$ :

$$\Gamma_h(h) = \bar{B}\left(1 + \frac{3}{2e^{3\bar{B}t(h)} + 1}\right) \quad (18)$$

where $t(h)$ may be derived from

$$h(t) = n_e(t) = \frac{h_0}{3}\left(2e^{-\bar{B}t} + e^{-4\bar{B}t}\right) \quad (19)$$

In principle, the function $\Gamma_h(h)$ may be obtained also in a closed-form (as solution of a quartic equation). The dependence (18) is written in Figure 6 as the dashed line. The solution is close to the data representing $\Gamma_h$ for $r = 1$. These analytical calculations confirm the influence of excited charge carriers distribution on TSFs. To obtain theoretically $\Gamma_h(h)$ for high retrapping coefficients it is necessary to model the kinetic processes more precisely on the basis of physical arguments. This modelling (including also presented examples as special cases) requires more calculations and will be presented in a separate paper.

## VI. CONCLUSIONS

The concept of TSFs allows for analytical formulation of trapping and recombination kinetics in the IC model. Therefore, it makes possible the analysis of TL in complex spatially correlated systems of traps and RCs which presumably are the norm in TL solid state detectors[4]. Present calculations confirm excellent invariance of TSFs with respect to thermal treatment (variable heating rates) and detrapping mechanism. These properties allow for simple numerical determination of TSFs. Example analytical calculations were performed also for a small two traps - two RCs cluster system. These calculations confirm the influence of excited charge carriers distribution on the determined TSFs. The future studies will consider the influence of initial filling on TSFs and the possibility of experimental determination of TSFs.

## ACKNOWLEDGEMENTS

This work was supported by the Polish State Committee for Scientific Research grant no. 3T10C01326